# Assessment of 7T task fMRI value over 3T task fMRI

Dalton H Bermudez[1]
[1]Department of Medical Physics, University of Wisconsin-Madison, WI

**Abstract**

Ultra-high field 7T fMRI offers notable advantages over 3T fMRI, including higher signal-to-noise and contrast-to-noise ratios, enabling finer spatial and temporal resolution. This study explores the differences in activation maps from Human Connectome Project datasets between 7T and 3T field strengths, focusing on visual identified using the Glasser atlas. Functional tasks for each scanner were designed to include visual stimuli, with data processed uniformly to ensure comparability. Results showed significantly higher beta coefficients for common regions of activation, such as V3A and V3B, in 7T compared to 3T datasets. This suggests that 7T fMRI data more accurately reflect the idealized time course of task-related conditions, likely due to improved sensitivity to blood oxygenation level-dependent (BOLD) signals. However, variations in experimental design and acquisition parameters between scanners complicate the direct comparison of beta coefficients.

**Introduction**

Functional magnetic resonance imaging (fMRI) at ultra-high field strengths, such as 7 Tesla (7T), has revolutionized the capabilities of neuroimaging research by providing significant advancements over the widely used 3 Tesla (3T) scanners. The primary advantage of 7T fMRI lies in its ability to achieve higher signal-to-noise ratio (SNR) and contrast-to-noise ratio (CNR), which result in enhanced spatial and temporal resolution (Ugurbil et al., 2013). These improvements are critical for investigating fine-scale neural structures and functional processes that are not readily detectable at lower field strengths. For example, 7T enables layer-specific imaging of the cortex, allowing researchers to probe the laminar organization of brain regions and their distinct roles in cognitive and sensory functions (Polimeni et al., 2010).

The higher sensitivity of 7T fMRI to blood oxygenation level-dependent (BOLD) signals further amplifies its utility for functional brain mapping. BOLD sensitivity increases approximately linearly with magnetic field strength, making 7T particularly well-suited for detecting subtle changes in neural activity (Triantafyllou et al., 2005). This advantage has significant implications for studying small and intricate structures such as the hippocampus, amygdala, and thalamic nuclei, which often exhibit weak BOLD responses at 3T (Olman et al., 2010). Additionally, 7T fMRI provides improved delineation of functional networks by reducing partial volume effects and enhancing spatial precision, thereby offering more accurate connectivity analyses (Heidemann et al., 2012).

Despite its numerous benefits, the use of 7T also poses challenges, including increased susceptibility artifacts, higher energy deposition, and cost considerations. However, ongoing advancements in acquisition protocols, such as parallel imaging and reduced field-of-view techniques, have mitigated many of these limitations (Okada et al., 2022). These developments have made 7T increasingly accessible and applicable in a range of neuroscience and clinical studies, spurring interest in systematically comparing its capabilities to those of 3T scanners.

This study aims to contribute to this growing body of knowledge by assessing the differences in activation maps derived from 7T and 3T Human Connectome Project (HCP) datasets. By leveraging datasets from the Human Connectome website, the analysis seeks to provide insights into the relative strengths and limitations of these two field strengths in functional neuroimaging.

*Acquisition Parameters*

Functional MRI (fMRI) data were acquired using both a 3T Siemens Skyra and a 7T Siemens Magnetom MR scanner, each equipped with advanced head coils and multi-band acceleration.

For the 3T protocol, data acquisition utilized a 32-channel head coil with a repetition time (TR) of 720 ms and an echo time (TE) of 33.1 ms. The flip angle was set at 52°, with a bandwidth (BW) of 2290 Hz/Px. The field of view (FOV) was 208 × 180 mm, covering 72 slices with 2.0 mm isotropic voxel resolution. To reduce acquisition time, a multi-band acceleration factor of 8 was employed.

For the 7T protocol, a Nova 32-channel Siemens head coil was used in conjunction with gradient-echo planar imaging (EPI). The TR was 1000 ms and the TE was 22.2 ms, with a flip angle of 45° and BW of 1924 Hz/Px. The in-plane FOV was 208 × 208 mm, covering 85 slices with a finer resolution of 1.6 mm isotropic voxels. A multi-band acceleration factor of 2 was utilized to maintain efficient temporal resolution while accommodating the higher spatial resolution of the 7T system.

*fMRI task*

The fMRI task for the 3T acquisition involved participants perception of emotions in face visual stimuli. The visual stimulus was interleaved between face expressions depicting different emotion and geometric visual stimuli. The face visual stimuli had a duration of 2 seconds, while the geometric figures had a duration of 1 second (Barch, et al. 2013). At the start of each emotion based 3T task experiment there was 8 seconds wait period before the first visual stimuli was presented to each subject. At the end of each block there was a task cue that had a duration of 3 seconds.

The fMRI task for the 7T acquisition involved presenting the subjects with interleaved rest periods and periods with natural video stimulus. The natural video stimulus had simultaneous auditory and visual stimulus. The corresponding duration of each rest and movie stimulus is depicted in Figure 1B. Despite the 7T movie task had more visual stimulus that differed from the 3T emotion perception task there were some frames of each movie trial that depicted different people face expressions. These two tasks between 7T and 3T for each corresponding subject were selected due to the relative similarity between stimuli.

*fMRI post-processing Methods*

The post-processing pipeline was kept consistent across both 3T and 7T datasets to ensure comparability. Structural T1-weighted (T1w) images underwent skull stripping and echo planar imaging (EPI) datasets were masked to exclude activations outside the brain. All TRs were in steady state, so no initial TRs were discarded. The corresponding motion regressors for each individual subjects' experiments were computed (Figure 2) and regressed out for the subject level analysis of each experiment of each individual subjects. Both T1w structural and EPI datasets were registered to the MNI anatomical template for standardization. A square wave BLOCK function, representing the stimulus onset and duration times with amplitude 1, was used for regression analysis. Motion artifacts exceeding 0.5 mm were censored from subject-level analysis, and outliers with values greater than 0.1 were excluded.

Post-processing was conducted using AFNI for volumetric analysis and SUMA for surface-based visualization, with identical parameters applied to both volumetric and surface-level subject fMRI data. This consistency ensures robust comparisons across modalities and facilitates accurate visualization of cortical activations.

**Results:**

*3T fMRI*

The volumetric activations maps for the 3T emotion task shows positive activations correlated with idealized stimulus timeseries in the visual cortex regions. The volumetric activation maps also show negative correlations with idealized stimulus timeseries in regions like the amygdala and the Thalamo-prefrontal network (Figure 3A). The positive activation in the visual cortex corresponds in the engagement of the visual system at perceiving a visual stimulus (Pfeifer et al 2019, Ikuko et al 2007, Mohamed et al. 2002).

Functional MRI (fMRI) studies have investigated the neural mechanisms underlying emotion perception, focusing on regions such as the amygdala and the thalamo-prefrontal network. While positive activations in these areas are commonly associated with emotional processing, certain tasks reveal negative activations, indicating complex neural dynamics. For instance, a study by Urry et al. (2006) examined the regulation of negative affect through cognitive reappraisal and found that increased activation in the prefrontal cortex was associated with decreased activation in the amygdala, suggesting an inhibitory relationship during emotion regulation tasks. Similarly, research by Phan et al. (2005) explored the neural substrates of voluntary emotion regulation using reappraisal and observed that successful downregulation of negative emotions corresponded with reduced amygdala activity and increased prefrontal cortex activation, highlighting the interplay between these regions in modulating emotional responses. In another study, Etkin et al. (2011) investigated implicit regulation of emotional conflict and reported that engagement of the rostral anterior cingulate cortex led to decreased amygdala activation, indicating that certain prefrontal regions can modulate amygdala responses even without explicit emotion regulation strategies. Furthermore, research by Ochsner et al. (2004) demonstrated that cognitive control mechanisms involving the dorsolateral prefrontal cortex can attenuate amygdala responses during the reappraisal of negative stimuli, underscoring the role of prefrontal regions in downregulating emotional reactivity. These studies collectively suggest that negative activations in the amygdala and thalamo-prefrontal network during emotion perception tasks reflect the engagement of regulatory processes aimed at modulating emotional responses. The dynamic interactions between these brain regions are crucial for understanding the neural basis of emotion regulation.

For the surface analysis of the 3T, cortical Glasser cortical parcels that had most significant activation were: V3A, V3B, V3, V1, 5R, 24dv, SFR, 24dd, SCEF, 8BR, 8Ad, 46, 46d, and AIP (Figure 3A). Most of the regions involved in positive activation were the visual cortical parcels.

*7T fMRI*

For the surface analysis of the 7T, cortical Glasser cortical parcels that had most significant activation were in the Pfop, PF, PFm, V3A, and V3B parcels of the cortex. (Figure 3B). Most of the regions involved the positive activation of the visual cortical parcels and parietal regions near the auditory cortex. The activation of visual cortical parcels along with activations in parietal parcel cortical surface regions have been previously observer for auditory/visual fMRI task. Glasser et al. (2016) reported activations in parietal regions during various tasks. Specifically, visual tasks elicited responses in areas such as the intraparietal sulcus (IPS), which is implicated in visual attention and processing. Auditory tasks, on the other hand, showed activations in regions like the superior parietal lobule (SPL), suggesting its role in auditory spatial attention.

Further research by Assem et al. (2020) examined the multiple-demand (MD) network using the HCP-MMP atlas. They identified a core set of 10 regions per hemisphere, including parietal areas, that are activated across diverse cognitive tasks, encompassing both auditory and visual modalities. This underscores the involvement of parietal regions in domain-general cognitive functions.

Since, the experimental task was different for both the 3T and 7T datasets, to compare the relative activation between each of the field strengths the study only focused on comparing Glasser cortical parcels common to both 3T and 7T datasets from the same subject. In this case, we only focused at comparing the relative activations in the Glasser cortical parcels in the V3A and V3B regions. To compare the relative activations from the surface region of interest (ROI) from the V3A and V3B, parcels were created on the surface renderings form the 3T and 7T T1w images. Then, the ROI of these two parcels regions were converted to surface files using AFNI's ROI2dataset functions. Those surface files were projected from surface to volume space using AFNI's 2dSurf2Vol functions. Finally, the corresponding beta coefficients from both the 3T and 7T regression analysis were extracted using AFNI's 2dmaskdump functions for only the mask corresponding to the V3A and V3B regions for both 3T and 7T datasets (Figure 4). A Wilcoxon rank sum test was performed between the beta coefficients of both these regions extracted from

3T and 7T datasets. The test showed a statistically significant difference in the beta coefficients extracted from both V3A and V3B regions of 7T compared to the beta coefficients of those same regions in 3T (Figure 5). Were the average beta coefficient of the 7T was 0.5 with a standard error of 0.02, while the average beta coefficient in the 3T was 0.29 with a standard error of 0.006.

**Conclusion**

Based on the results of this study, the analysis highlights that while it is challenging to isolate the effects of beta coefficients due to differences in acquisition protocols and experimental task design, preliminary findings suggest that 7T fMRI data demonstrates significantly higher beta coefficients for parcel regions with comparable significant activation compared to that of the 3T data. This indicates that 7T fMRI may enable more precise tracking of parcel activations that align more closely with the idealized time courses of the task conditions compared to 3T fMRI.

Future studies should aim to address the variability introduced by differences in scanner parameters and acquisition protocols. Achieving a direct comparison between 3T and 7T data would benefit from employing consistent experimental designs and acquisition parameters across field strengths. Additionally, group-level statistical analysis of activation maps should be considered in future work to derive more robust and generalizable results. Incorporating this approach would further validate findings and enhance the understanding of the differences in neural activation patterns between 3T and 7T fMRI, contributing to advancements in functional neuroimaging.

**Acknowledgements**

ChatGPT was used to provide concrete scientific literature that supported the results and for grammar corrections.

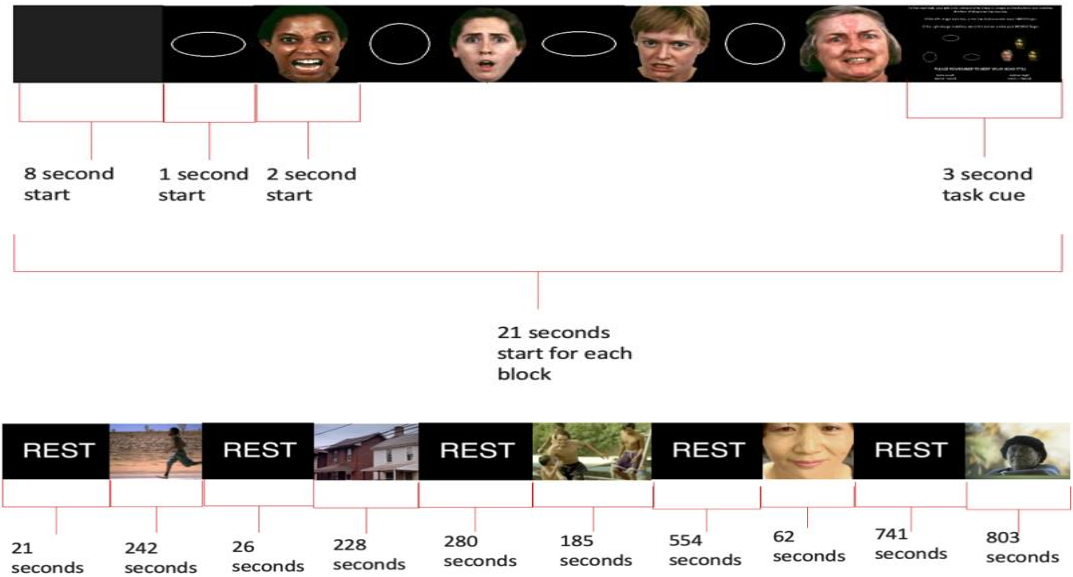

**Figure 1 A)** The stimulus EMOTION experiment performed on subjects for 3T task EPI fMRI scan. **B)** The stimulus MOVIE task performed on subjects for 7T task EPI fMRI scan.

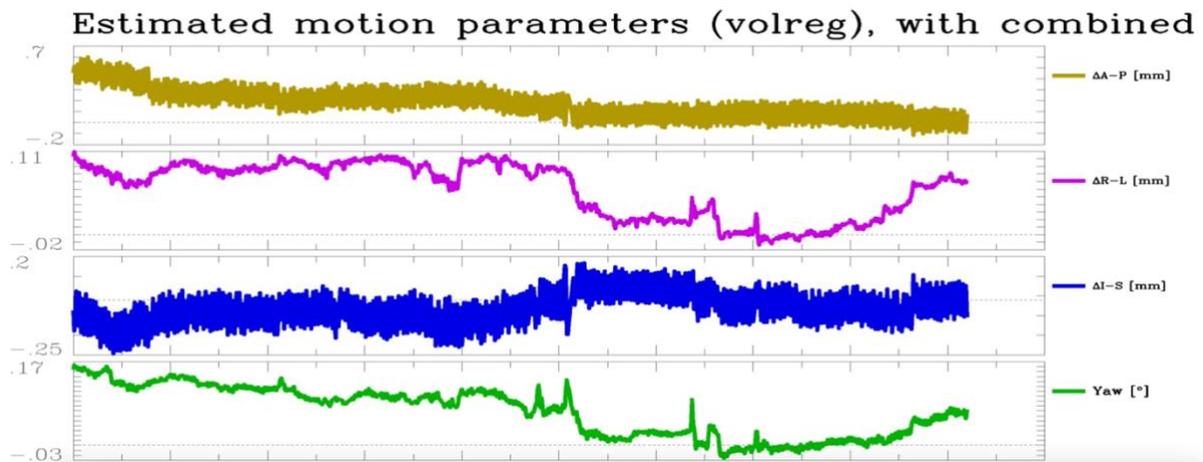

**Figure 2** shows an example of the motion based regressor computed and regressed out of the subject based analysis for the 7T MOVIE task experiment.

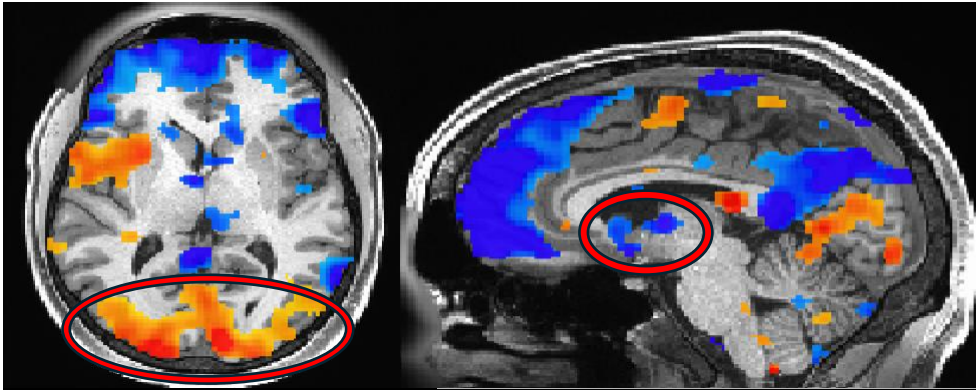
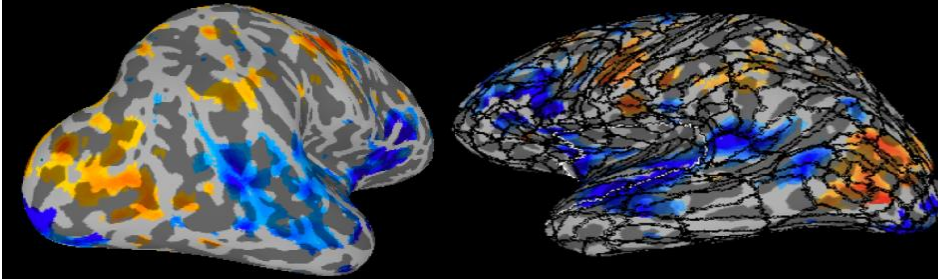
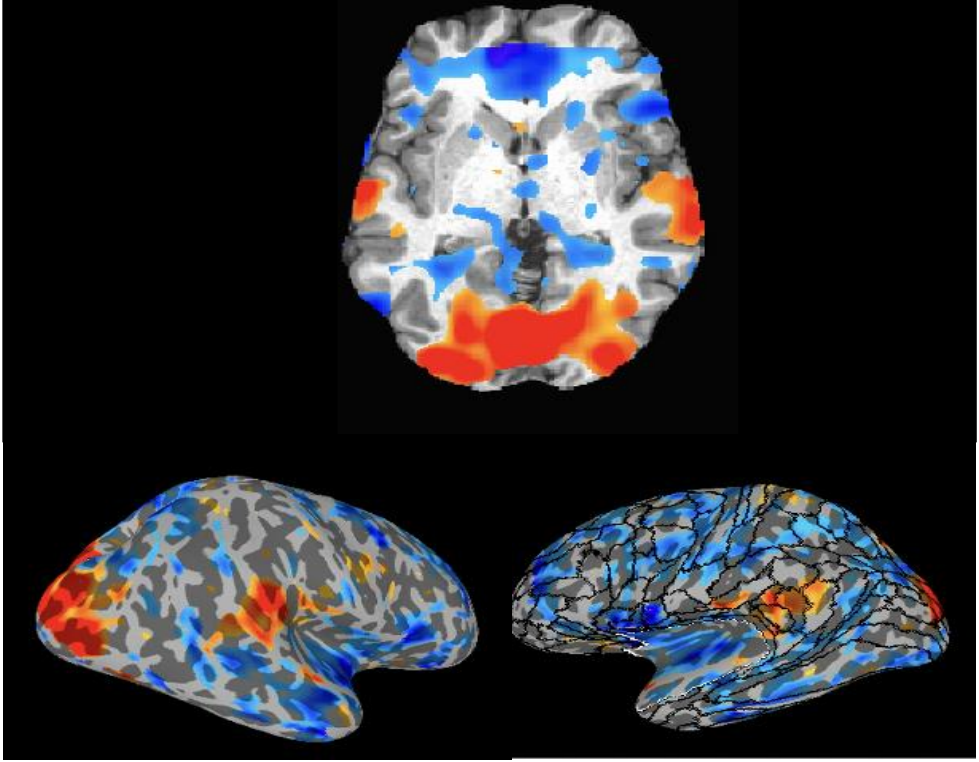

**Figure 3: A)** Shows the volumetric, surface, and surface with Glasser Parcels overlayed with statistical mapping of most significant activations regions for 3T. **B)** Shows the volumetric, surface, and surface with Glasser Parcels overlayed with statistical mapping of most significant activation regions for 7T. Both **A)** and **B)** corresponds to the same subject.

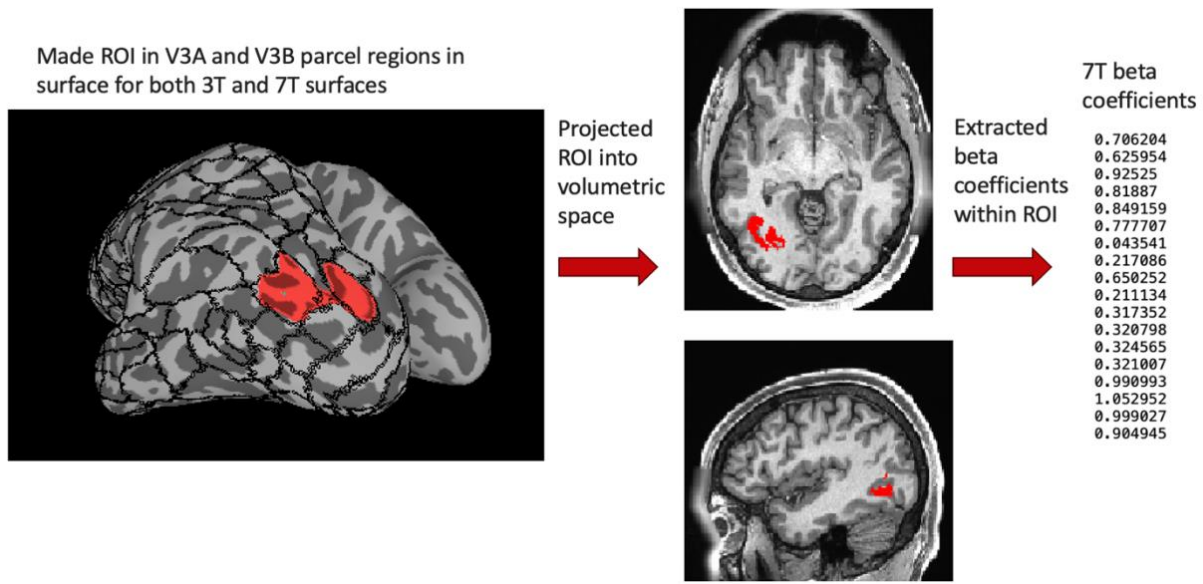

**Figure 4:** Shows the pipeline to extract the beta coefficients for both V3A and V3B parcels which share common significant positive activations for both 3T and 7T scans for the same subject. ROI were made to mask V3A and V3B Glasser parcel regions in the surfaces. The ROI were then projected onto the volume and the volume mask of these regions were used to extract beta coefficients in V3A and V3B regions for both 3T and 7T datasets.

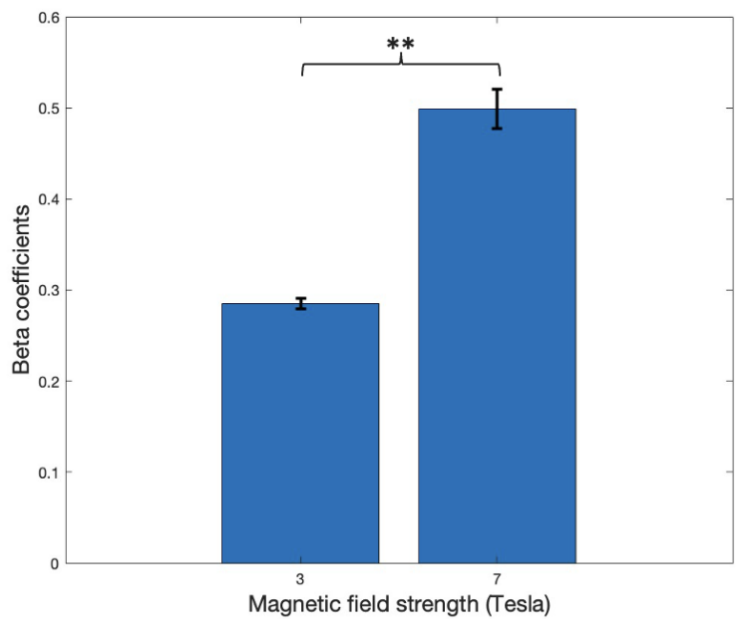

**Figure 5:** Bar graph shows the distribution of beta coefficients for both common activated regions of V3A and V3B regions in 3T and 7T datasets for the same subject. A Wilcoxon rank sum test was performed between the beta coefficients of both these regions extracted from 3T and 7T datasets. The test shows that there is a statistically significant difference in the beta coefficients for both of this Glasser parcel regions between 3T and 7T datasets.

**References:**

[1] Uğurbil K, Xu J, Auerbach EJ, Moeller S, Vu AT, Duarte-Carvajalino JM, Lenglet C, Wu X, Schmitter S, Van de Moortele PF, Strupp J, Sapiro G, De Martino F, Wang D, Harel N, Garwood M, Chen L, Feinberg DA, Smith SM, Miller KL, Sotiropoulos SN, Jbabdi S, Andersson JL, Behrens TE, Glasser MF, Van Essen DC, Yacoub E; WU-Minn HCP Consortium. Pushing spatial and temporal resolution for functional and diffusion MRI in the Human Connectome Project. Neuroimage. 2013 Oct 15;80:80-104. doi: 10.1016/j.neuroimage.2013.05.012. Epub 2013 May 21. PMID: 23702417; PMCID: PMC3740184.

[2] Polimeni JR, Fischl B, Greve DN, Wald LL. Laminar analysis of 7T BOLD using an imposed spatial activation pattern in human V1. Neuroimage. 2010 Oct 1;52(4):1334-46. doi: 10.1016/j.neuroimage.2010.05.005. Epub 2010 May 9. PMID: 20460157; PMCID: PMC3130346.

[3] Triantafyllou C, Polimeni JR, Wald LL. Physiological noise and signal-to-noise ratio in fMRI with multi-channel array coils. Neuroimage. 2011 Mar 15;55(2):597-606. doi: 10.1016/j.neuroimage.2010.11.084. Epub 2010 Dec 16. PMID: 21167946; PMCID: PMC3039683.

[4] Olman CA, Davachi L, Inati S. Distortion and signal loss in medial temporal lobe. PLoS One. 2009 Dec 3;4(12):e8160. doi: 10.1371/journal.pone.0008160. PMID: 19997633; PMCID: PMC2780716.

[5] Heidemann RM, Ivanov D, Trampel R, Fasano F, Meyer H, Pfeuffer J, Turner R. Isotropic submillimeter fMRI in the human brain at 7 T: combining reduced field-of-view imaging and partially parallel acquisitions. Magn Reson Med. 2012 Nov;68(5):1506-16. doi: 10.1002/mrm.24156. Epub 2012 Jan 9. PMID: 22231859.

[6] Okada T, Fujimoto K, Fushimi Y, Akasaka T, Thuy DHD, Shima A, Sawamoto N, Oishi N, Zhang Z, Funaki T, Nakamoto Y, Murai T, Miyamoto S, Takahashi R, Isa T. Neuroimaging at 7 Tesla: a pictorial narrative review. Quant Imaging Med Surg. 2022 Jun;12(6):3406-3435. doi: 10.21037/qims-21-969. PMID: 35655840; PMCID: PMC9131333.

[7] Deanna M. Barch, Gregory C. Burgess, Michael P. Harms, Steven E. Petersen, Bradley L. Schlaggar, Maurizio Corbetta, Matthew F. Glasser, Sandra Curtiss, Sachin Dixit, Cindy Feldt, Dan Nolan, Edward Bryant, Tucker Hartley, Owen Footer, James M. Bjork, Russ Poldrack, Steve Smith, Heidi Johansen-Berg, Abraham Z. Snyder, David C. Van Essen, Function in the human connectome: Task-fMRI and individual differences in behavior, NeuroImage, Volume 80, 2013, Pages 169-189, ISSN 1053-8119, https://doi.org/10.1016/j.neuroimage.2013.05.033.

[8] Pfeifer G, Ward J and Sigala N (2019) Reduced Visual and Frontal Cortex Activation During Visual Working Memory in Grapheme-Color Synaesthetes Relative to Young and Older Adults. *Front. Syst. Neurosci.* 13:29. doi: 10.3389/fnsys.2019.00029

[9] Ikuko Mukai, David Kim, Masaki Fukunaga, Shruti Japee, Sean Marrett, Leslie G. Ungerleider. Activations in Visual and Attention-Related Areas Predict and Correlate with the Degree of Perceptual Learning. Journal of Neuroscience 17 October 2007, 27 (42) 11401-11411; **DOI:**10.1523/JNEUROSCI.3002-07.2007

[10] Mohamed FB, Pinus AB, Faro SH, Patel D, Tracy JI. BOLD fMRI of the visual cortex: quantitative responses measured with a graded stimulus at 1.5 Tesla. J Magn Reson Imaging. 2002 Aug;16(2):128-36. doi: 10.1002/jmri.10155. PMID: 12203759.

[11] Urry HL, van Reekum CM, Johnstone T, Kalin NH, Thurow ME, Schaefer HS, Jackson CA, Frye CJ, Greischar LL, Alexander AL, Davidson RJ. Amygdala and ventromedial prefrontal cortex are inversely coupled during regulation of negative affect and predict the diurnal pattern of cortisol

secretion among older adults. J Neurosci. 2006 Apr 19;26(16):4415-25. doi: 10.1523/JNEUROSCI.3215-05.2006. PMID: 16624961; PMCID: PMC6673990.

[12] Phan KL, Fitzgerald DA, Nathan PJ, Moore GJ, Uhde TW, Tancer ME. Neural substrates for voluntary suppression of negative affect: a functional magnetic resonance imaging study. Biol Psychiatry. 2005 Feb 1;57(3):210-9. doi: 10.1016/j.biopsych.2004.10.030. PMID: 15691521.

[13] Etkin A, Egner T, Kalisch R. Emotional processing in anterior cingulate and medial prefrontal cortex. Trends Cogn Sci. 2011 Feb;15(2):85-93. doi: 10.1016/j.tics.2010.11.004. Epub 2010 Dec 16. PMID: 21167765; PMCID: PMC3035157.

[14] Ochsner KN, Ray RD, Cooper JC, Robertson ER, Chopra S, Gabrieli JD, Gross JJ. For better or for worse: neural systems supporting the cognitive down- and up-regulation of negative emotion. Neuroimage. 2004 Oct;23(2):483-99. doi: 10.1016/j.neuroimage.2004.06.030. PMID: 15488398.

[15] Glasser, M., Coalson, T., Robinson, E. *et al.* A multi-modal parcellation of human cerebral cortex. *Nature* **536**, 171–178 (2016). https://doi.org/10.1038/nature18933

[16] Moataz Assem, Matthew F Glasser, David C Van Essen, John Duncan, A Domain-General Cognitive Core Defined in Multimodally Parcellated Human Cortex, *Cerebral Cortex*, Volume 30, Issue 8, August 2020, Pages 4361–4380, https://doi.org/10.1093/cercor/bhaa023